\documentclass
[prl,showpacs,
tightenlines,
nobibnotes,aps,prl,
secnumarabic,
amsmath,amssymb,floatfix]{revtex4}
\usepackage{docs}
\renewcommand{\subsection}[1]{~\newline~{\bf{#1:}}}
\pacs{87.10.+e, 89.75.Fb, 87.16.Ac, 87.23.Kg}

\newif\ifpdf
\ifx\pdfoutput\undefined
\pdffalse 
\else
\pdfoutput=1 
\pdftrue
\fi

\ifpdf
\usepackage[pdftex]{graphicx}
\else
\usepackage{graphicx}
\fi

\begin{document} 
\title{Systematic identification of statistically significant network measures} 
\author{Etay Ziv,$^{1,2}$ Robin Koytcheff,$^3$ Manuel Middendorf,$^4$Chris Wiggins$^{3,5}$}
\affiliation{$^1$College of Physicians and Surgeons,\\
$^2$Department of Biomedical Engineering,\\ 
$^3$Department of Applied Physics and Applied Mathematics,\\
$^4$Department of Physics,\\
$^5$Center for Computational Biology and Bioinformatics,\\ 
Columbia University, New York NY 10027}
\begin{abstract} 
We present a novel graph embedding space (i.e., a set of measures on graphs) for performing statistical analyses of networks. Key improvements over existing approaches include discovery of ``motif-hubs" (multiple overlapping significant subgraphs), computational efficiency relative to subgraph census, and flexibility (the method is easily generalizable to weighted and signed graphs). The embedding space is based on 
{\it scalars}, functionals of the adjacency matrix representing the network. {\it Scalars} are global, involving all nodes; although they can be related to subgraph enumeration, there is not a one-to-one mapping between scalars and subgraphs. Improvements in network randomization and significance testing--we learn the distribution rather than assuming gaussianity--are also presented.
The resulting algorithm establishes a systematic approach to the identification of the most significant scalars and suggests machine-learning techniques for network classification. 

\end{abstract}
\maketitle
\begin{figure}[tbh]
 \centerline{\includegraphics[width=3.5in] {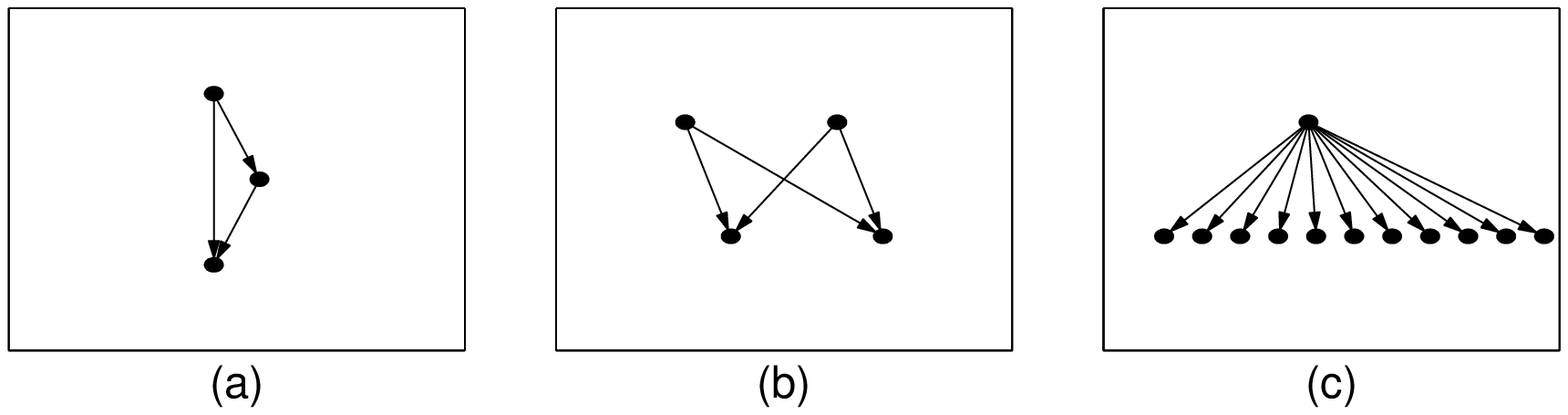}}
  \caption{Three structures recovered after hard-localization on significant scalars in \emph{E. coli} validates our method. Note that these three structures were identified as statistically significant using one unique, systematic enumeration of scalars. (a)$\cal{T}$$_{030}$ (Fig 1a in \cite{Alon1}) subgraph contributing to the scalar $\sum(DA^TAA)$. (b), (Fig 1e in \cite{Alon1}), subgraph contributing to the scalar $\sum(DA^TAUA^TA)$. (c), (Fig 1c in \cite{Alon1}), subgraph contributing to the scalar, $\sum(DA^TAA^TA)$.}
  \label{fflsimdor} \end{figure}

\begin{figure}[tbh]
  \centerline{\includegraphics[width=3in] {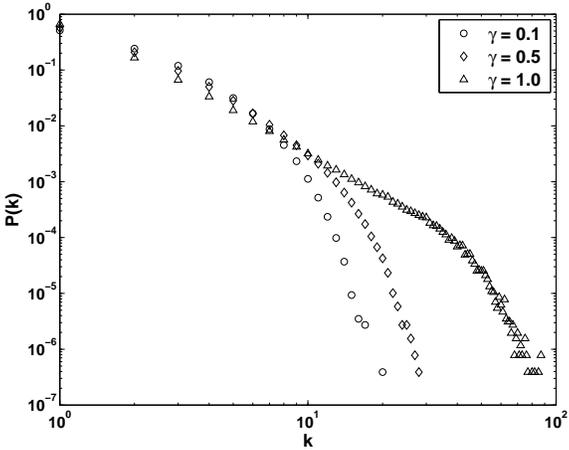}}
  \caption{Degree distributions of networks generated using the Barabasi and Albert preferential attachment model with the tunable parameter, $\gamma$, such that $A_{k}=k^{\gamma}$, where $A_{k}$ is the probability of a new vertex attaching to an existing vertex with $k$ links. All of these networks are the same size and have the same density ($423$ nodes, $519$ edges), but differ in their degree distributions.}
  \label{alphs_coli} \end{figure}

\begin{figure}[tbh]
    \centerline{\includegraphics[width=3in] {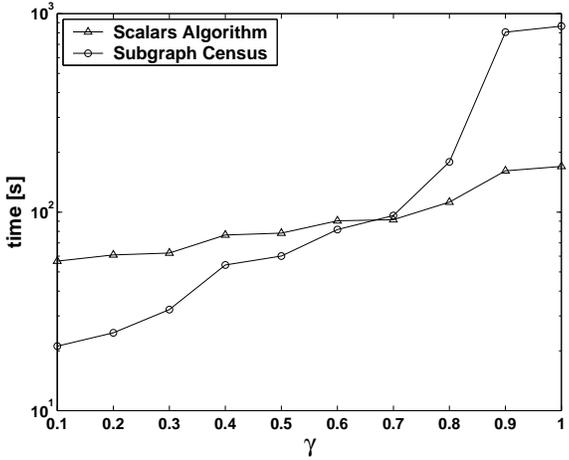}}
  \caption{A numerical experiment comparing efficiency of ``traditional" subgraph counting algorithm (green) and the proposed ``scalars" algorithm (red), as a function of $\gamma$, a parameter which tunes the degree of scale-invariance in the network (see Figure \ref{alphs_coli}). The number of nodes and the density in the networks were kept constant and equal to those of the \emph{E. coli} network tested in the manuscript. Scale-free properties similar to naturally-occurring networks emerge with linear preferential attachment, where $\gamma=1$ (e.g., at $\gamma \sim 1$ the network contains hubs whose degree is similar to the degree of hubs in the \emph{E. coli} network). We see here, the scalars algorithm becomes more efficient at $\gamma>0.7$.}
  \label{wvw} \end{figure}

\begin{figure}[tbh]
  \centerline{\includegraphics[width=2.5in] {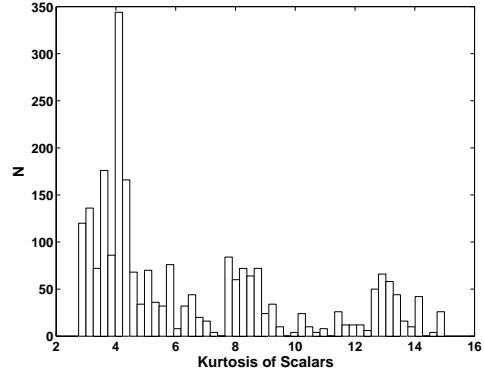}}
  \caption{A histogram of the kurtosis (a measure of the degree of peakedness of a distribution) for scalars demonstrates many non-Gaussian distributions (i.e., distributions with kurtosis greater than or less than $3$). This is also the case for subgraph distributions and hence we employ density estimation rather than assume Gaussianity.}
  \label{kurt} \end{figure}

\begin{figure}[tbh]
    \centerline{\includegraphics[width=2.5in] {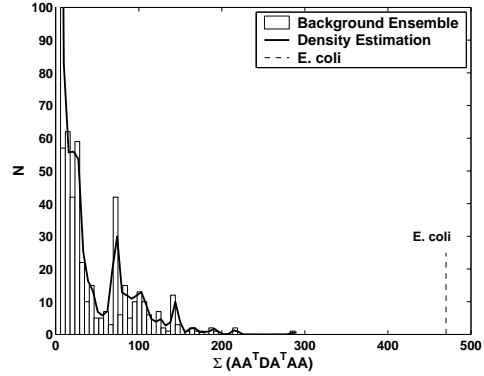}}
  \caption{The scalar $\sum(AA^TDA^TAA)$ has a value of $470$ in \emph{E. coli}. Kernel density estimation of the distribution obtained from this scalar for networks generated from the randomization yields a log-likelihood of $\log(p)<-708$ for this scalar. See \ref{soft1} and \ref{hard1} for soft- and hard- localizations of this scalar, respectively.}
  \label{kde1} \end{figure}

\begin{figure}[tbh]
    \centerline{\includegraphics[width=2.5in] {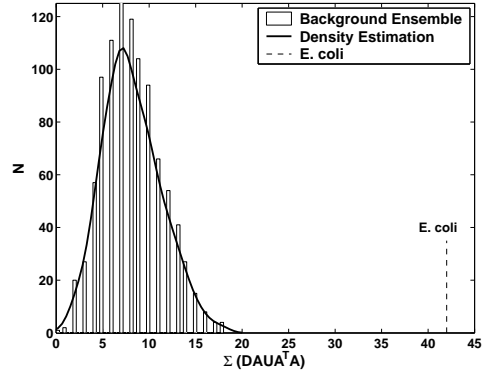}}
  \caption{The scalar $\sum(DAUA^TA)$ has a value of $42$ in \emph{E. coli}. Kernel density estimation of the distribution obtained from this scalar for networks generated from the randomization yields a log-likelihood of $\log(p)=-525$ for this scalar. Soft- and hard-localizations yield a feed-forward topology.}
  \label{kde2} \end{figure}

\begin{figure}[tbh]
    \centerline{\includegraphics[width=2.5in] {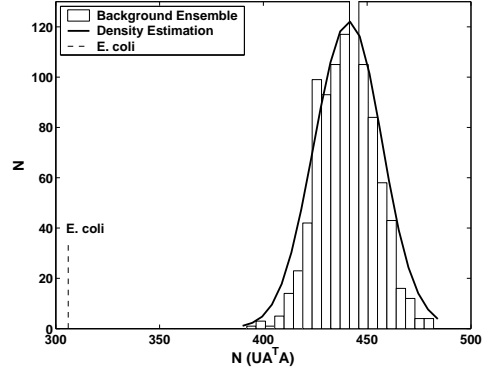}}
  \caption{The scalar $\cal {N}$$(UA^TA)$ has a value of $306$ in \emph{E. coli}. Kernel density estimation of the distribution obtained from this scalar for networks generated from the randomization yields a log-likelihood of $\log(p)=-163$. In \emph{E. coli} this scalar is significantly underrepresented.  The walk that it counts, namely moving forward, and then backward, but not ending up at the starting point, emphasizes a fan-in topology. These fan-in structures are thus not well-represented in \emph{E. coli}, a finding which supports work in the computational biology literature in which such prior assumptions about the network structure are used to infer genetic interactions \cite{Husmeier}.}
  \label{kde3} \end{figure}

\begin{figure}[tbh]
  \centerline{\includegraphics[width=2.5in] {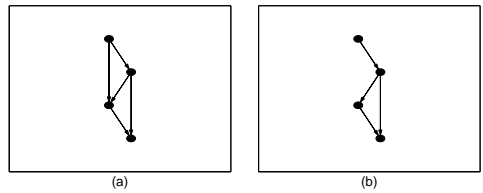}}
  \caption{In \emph{E. coli}, soft-localization of the most significant scalar, $\sum(AA^TDA^TAA)$, yields   these two representative subgraphs at equal relative fractions, which we call (a) ``+FFL", feed-forward loop with an input and (b) ``FFB", feed-forward box.}
  \label{soft1} \end{figure}

\begin{figure}[tbh]
  \centerline{\includegraphics[width=2.5in] {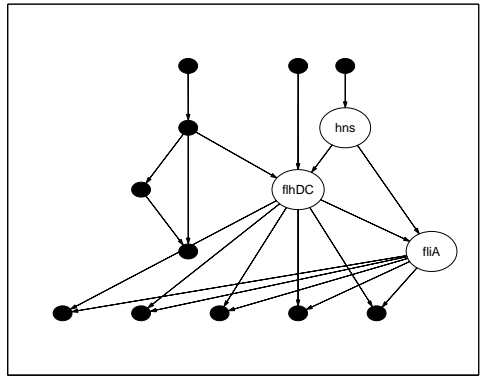}}
  \caption{In \emph{E. coli} hard-localization of the most significant scalar, $\sum(AA^TDA^TAA)$, yields this $14$ node topology. Note the presence of ``motif-hubs" -- statistically significant subgraphs which share one or more nodes. For example, there are five overlapping feed-forward boxes which share three common genes arranged in a feed-forward loop, $hns$, $fliA$, and $flhDC$. These three genes act as transcription regulators for the \emph{E. coli} flagellar pathway. $Hns$ and $crp$ mutants are nonmotile, but overexpression of the ``master operon" $flhDC$ restores, in part, motility in these mutant strains \cite{Soutourina}}. 
  \label{hard1} \end{figure}

\begin{figure}[tbh]
    \centerline{\includegraphics[width=2.5in] {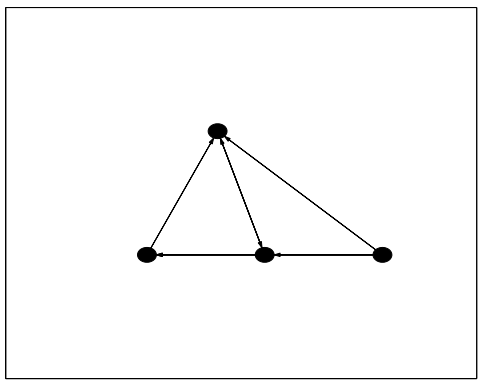}}
  \caption{In \emph{S. cerevisiae}, both hard- and soft-localization of the significant scalar $\sum(DAAA^TDAA)$ yields this densely clustered $4$ node topology which includes a mutual dyad and a $3$-cycle. Unlike the \emph{E. coli} network, this network contains feedback interactions.}
  \label{m3c} \end{figure}

\begin{figure}[tbh]
  \centerline{\includegraphics[width=2.5in] {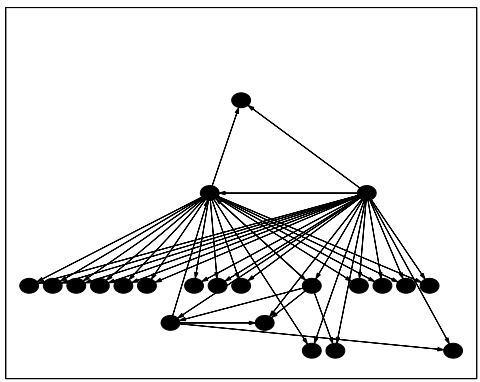}}
  \caption{In \emph{S. cerevisiae}, hard-localization of the significant scalar $\sum(DAAAUA^TA)$ yields this interesting $22$ node topology. Note again the fan-out structure whereby two genes regulate a very similar set of genes.}
  \label{yeast22} \end{figure}

\begin{figure}[tbh]
    \centerline{\includegraphics[width=2.5in] {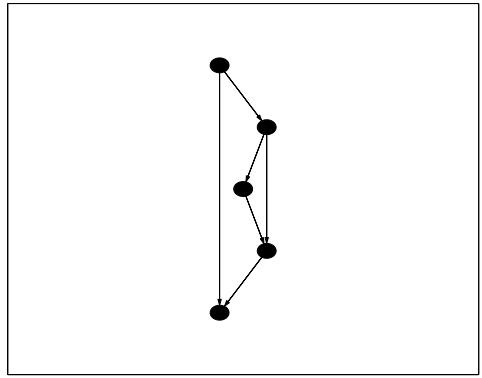}}
  \caption{In \emph{S. cerevisiae}, soft-localization of the significant scalar, $\sum(DAAAUA^TA)$ yields this $5$ node topology as the most representative subgraph. Interestingly, the structure can be seen as a hierarchical feed-forward loop. For example, if we replace the $3$ node feed-forward loop with an effective node, that node itself becomes the parent layer of another feed-forward loop.}
  \label{soft2} \end{figure}

 
\ifpdf
\DeclareGraphicsExtensions{.pdf, .jpg, .tif}
\else
\DeclareGraphicsExtensions{.eps, .jpg}
\fi
\renewcommand{\baselinestretch}{2.4}
\normalsize
{\bf{Background}:} 
Recent studies of real-world biological, social, and technological networks have catalyzed an explosion of research from a broad range of disciplines. Much of the effort in this emerging field has focused on characterizing the structure of networks using various statistical properties that are local (analysis relies on subset of nodes) or global
(relying on all nodes) in scope. The former analysis includes subgraph census (comparing frequency of subgraph occurrences in a given graph with those over a distribution of graphs \cite{HandL1, 
Alon1}), while examples of the latter include path lengths and degree distributions (see citations in \cite{newman}).\\  
To study local structure statistics, sociologists developed the $k$-subgraph census, an enumeration of all possible subgraphs of $k$ nodes 
appearing in networks. For example, sociologists used the 3-subgraph census, compared with 3-subgraph distributions in randomized graphs, to quantify
network transitivity \cite{WassermanFaust,HandL2, DandL1} (in the context of a social network, high transitivity means that many of your friends are friends with each other). Applying such techniques first to the \emph{E. coli} genetic network \cite{Alon1} and later to various biological and physical networks \cite{Alon2}, Milo et al showed that different networks have different ``most significant" subgraphs. \\
Major limitations of these subgraph approaches include computational cost and generalizability. 
The number of isomorphism
classes of {\em digraphs} grows rapidly with graph size \cite{Harary1, WassermanFaust} and 
subgraph isomorphism is an NP-complete problem \cite{NP1971}~\footnote{Digraphs (directed graphs, or graphs whose edges have directionality) are ismorphic if there exists a relabeling of their vertices such that the two graphs are identical. NP class consists of decision problems whose solution can be found in polynomial time on a non-deterministic Turing machine. NP-complete problems are a subset of NP which are particularly hard. Given $2$ graphs, $G_1$ and $G_2$, subgraph isomorphism asks, if $G_1$ is isomorphic to a subgraph of $G_2$. }. 
These computational limitations bias results, since structures with more than three or four nodes would not be counted. Moreover, it is not readily obvious how to extend subgraph census to weighted and/or signed graphs. This is particularly relevant for genetic regulatory networks in which the interactions can be described quantitatively via binding affinities and qualitatively as activating or repressing, or similarly neuronal networks, in which the interactions are often weighted by the number of synapses between neurons and can also exhibit excitatory and inhibitory behaviors. \\
In their groundbreaking work, Shen-Orr et al. identified three significant motifs in the \emph{E. coli} genetic network.
However, rather than counting all structures up to a given size, the authors had to resort to posing putative significant structures, thus making prior assumptions about which subgraphs are important. One topology was found by enumerating all $3$-node subgraphs in the network; a second by searching for single regulator genes regulating at least $13$ distinct operons; and the third by presenting a clustering algorithm based on several new parameters. 
Similarly, in \cite{Young} six different subgraphs were defined using six different algorithms. Rather than finding subsets of motifs via tailored, parameterized, and thresholded algorithms, a single, generalizable method for identifying motifs is needed. \\ 
In this manuscript, we first present a novel embedding space for networks, which is 
(i) computationally efficient as compared to subgraph census for naturally-occurring networks and (ii) easily applicable to weighted and signed graphs. We then employ this space in a single, generalizable algorithm to discover arbitrarily-large, statistically significant network measures in the \emph{E. coli} and the \emph{S. cerevisiae} genetic regulatory networks. New results are presented about the structure of these networks, including the presence of overlapping significant subgraphs. We also introduce a new randomization scheme which generates independent identically distributed samples rather than a markov chain, and we integrate density estimation into our significance testing rather than asserting Gaussianity.   \\
{\bf{Motivation}:} 
As a motivating example,
we consider the 
$3$-subgraph, $\cal{T}$$ _{030}$, defined as  the triad of nodes $i$, $j$, and $k$, and edges $i$$\rightarrow$$j$, $j$$\rightarrow$$k$, $i$$\rightarrow$$k$ (See \ref{fflsimdor}a), which we represent by its adjacency matrix, $A$ ($A_{ij} \equiv 1$ if the $j^{th}$ node is the parent of $i^{th}$ node, and 0 otherwise). 
This nomenclature references the earliest work in subgraph census \cite{HandL1, WassermanFaust}.
We observe that the number of $\cal {T}$$_{030}$ in this graph is trivially found by simple matrix manipulation of its adjacency matrix as follows. The 
trace of the square of $A$ multiplied by its transpose yields $1$. Indeed, a count of $\cal {T}$$_{030}$ subgraphs in $\it{any}$ graph can be obtained in this way.
Similarly, other subgraphs can be enumerated in terms of the adjacency matrix, $A$, its transpose, $A^T$, diagonal projection operator D and its complement U defined for any matrix Q by $[D(Q)]_{ij}=Q_{ij}\delta_{ij}$, and $U(Q)=Q-D(Q)$, respectively.  
Note that we do not use Einstein's summation convention and $D$ is not the trace.\\
$A,A^T,D,U$ can be visualized as motion on the digraph: $A$ or $A^T$ represent 
moving one step forward or backward, respectively; $D$ represents restriction to 
closed paths; $U$ represents open paths.
In terms of the adjacency matrix and the functions that act on it, we can enumerate $\cal{T}$$_{030}$ as $\sum(D(A^T(A^2)))$. Reading this expression from right to left, we start at a node, move two steps forward ($A^2$), then one step backward ($A^T$), and arrive at the original starting node ($D$). By summing all $n^2$ elements of the resulting matrix, we obtain a count of $\cal {T}$$_{030}$. Instead of summing, we could also count the number of nonzero elements, $\cal N$. 
These operations on the resulting matrix, $\sum$ or $\cal N$, yield the number of distinct paths between all pairs of endpoints or the number of distinct pairs of endpoints, respectively. \\
We define a \emph{word} to be the matrix built from the letters, $A$, $A^T$, $D$ and $U$, and a  \emph{scalar} as the integer obtained from the operations, $\sum$ or $\cal N$ on a word. An enumeration of words and subsequent evaluation of scalars allows us to embed a given network in an infinite-dimensional space.
To enumerate words we systematically combine letters. Obvious redundancies can be eliminated (\emph{e.g.} $U^2=U$, $D^2=D$, $UD=DU=0$). We construct words by combining letters such that each letter acts on {\emph{everything}} to its right. As an example, the word $D(A^T(A(A)))$ is constructed from the letter $D$ acting on $A^T$ acting on $A$ acting on $A$. The scalar is obtained by evaluating either $\sum$ (the sum over) or $\cal N$ (the number of nonzero elements in) the word. Other choices for construction of words are possible (e.g., using different combinations of parentheses, $D(A^T)*(A^2)$). Our method can easily generalize to include these words. For simplicity, herein we will assume parentheses are implicit and write words without the parentheses. Thus, $D(A^T(A(A)))$ will be written $DA^TAA$. \\
{\bf{Proposal}:}
Given this embedding space for networks, essentially a set of measures on a network, one can then employ standard tools from statistics and machine learning to characterize a network of interest.
For the specific application of identifying statistically significant features of a network, this tantalizing observation motivates the technique presented here:
\begin{enumerate}
\item
systematically enumerate words;
\item
evaluate the scalars obtained from these words for a graph of interest;
\item
compare scalars with the distribution obtained by evaluating scalars
over a randomly-generated distribution of matrices, 
thus finding statistically significant {\it scalars}.  
\end{enumerate} 
The fact that scalars are based on combinations of functionals of the adjacency matrix makes our method easily extendable to weighted and signed graphs. For example, in the former one could simply use the weight matrix in place of the adjacency matrix;  in the latter, one could use two adjacency matrices representing the two types of interactions. \\
As stated above, a major limitation of subgraph census is computational efficiency. Here we present analytic and numerical comparisons between subgraph census and our scalars technique. Traditional algorithms count subgraphs by performing walks \cite{Alon2, Barabasi}. Given a graph with $N$ nodes and $M$ edges, the computational cost of subgraph counting grows exponentially in the size of the subgraph, $n$, worse than exponentially in the density, $M/N$, and is traditionally infeasible for $n>4$, especially in scale-free networks \cite{Alon2, Barabasi, Mirny}. 
In scalar calculation, computational complexity is upper bounded by $N^3\sum_i{(\ell_i-1)}$, where $\ell_i$ is the number of letters in scalar $i$ and the sum is performed over all scalars. While complexity grows exponentially in the number of letters, the exponential term is independent of the density and the degree distribution. Thus feature selection using scalars is especially suited for dense, clustered, or scale-free networks. \\
This observation is particularly relevant as many naturally-occuring networks have heterogenous degree distributions \cite{BarabasiAlbert}. To quantify the effect of degree distributions on the performance of the two algorithms, we benchmark subgraph census against our scalars method using randomly-generated networks. 
We generate multiple graphs of the same size and density as the \emph{E. coli} genetic regulatory network, but with different degree distributions, using the class of growing random network (GRN) models with tunable parameter $\gamma$, first proposed by Krapivsky \emph{et al.} \cite{Krapivsky} as a generalization of the cumulative advantage or preferential attachment models \cite{Price, BarabasiAlbert}. In the GRN model, at every time step a new node is added, and with probability $A_{k}$, an edge is created between the new node and an existing node with $k$ edges, where $A_{k}=k^{\gamma}$. 
The preferential attachment parameter, $\gamma$ acts to tune the degree of heterogeneity in the degree distribution. As $\gamma$ approaches $1$, or linear preferential attachment, the degree distribution becomes more heavy-tailed, and thus more similar to naturally-occuring networks. 
In Figure \ref{alphs_coli} we show degree distributions for graphs generated at three different values of $\gamma$. 
In Figure \ref{wvw}, we demonstrate how the scalars method significantly improves computational time for these types of degree distributions, which many biological (as well as technological and sociological) networks evidence.\\
{\bf{Background Ensemble}:}  
A vast literature discusses different randomly generated network models
\cite{WassermanFaust, Katz, Sjniders1, Rao,
  Roberts, Molloy, Bender,newman}.  In \cite{Alon1} a random model was used
which preserved $N(k_{+}, k_{-})$: in-degree and out-degree of each node (random matching of a given in- and out-degree sequence is also known as the configuration model \cite{newman, Molloy}).		
This can
be done efficiently by representing the graph as ordered lists of parents 
and children. The
number of times the node occurs in the parent (child) list is the
node's out (in) degree. Permuting one of the two lists, one attains
the configuration model. Pathological permutations give rise to multiple edges and
self-interactions. Individual pathologies can be corrected
at little additional computational expense (see \texttt{fixpath} in \cite{code} for details).  
In this case, we
preserve $N(k_{+}, k_{-}, k_{0})$, the joint distribution for
in- and out-degree and self-interactions. 
In \cite{Maslov, Mirny} a similar ensemble is used where multiple edges are disallowed; however, our approach differs in the following respects: (i) our algorithm is a more efficient single shuffle rather than multiple swaps, (ii) iteratively rewiring requires the introduction of another cut-off parameter, defining how many rewiring steps are needed; shuffling obviates the need for this additional parameter, (iii) iterative swapping generates markov chain realizations whereas shuffling generates independent, identically distributed samples (iv) we preserve self-interactions.\\ 
{\bf{Statistical Significance}:}
In the past, statistical analyses of subgraphs have relied on $z$-scores or empirical sample estimates of probabilities. In Figure \ref{kurt} we show that many features (both for subgraphs and for scalars) are not gaussian, so $z$-scores are inappropriate measures of deviation from the background ensemble. Empirical sample estimates are also problematic, for example, if the distribution is under-sampled. Instead we apply standard tools from machine learning, namely \emph{kernel density estimation} and \emph{cross-validation} to learn the distribution from the sample data. Cross-validation is a model evaluation method where 
model learning relies on part of the data, while model testing 
relies on the rest of the data, the holdout set. K-fold 
cross-validation repeats the holdout method $k$ times.
To quantify a network's deviation from the background ensemble, we learn the distribution for each scalar and measure deviation as the likelihood that an observation was drawn from the background distribution.
Given a graph and our model, we 
collect $m$ realizations
and estimate the probability density $p(W_j=w)$ for a scalar $j$ to have a value $w$ using 
Gaussian kernel density estimation~\cite{Hastie}:
\begin{eqnarray}
p_{\lambda_*}(w)=\frac{1}{m}\sum_{i=1}^m{\frac{e^{-\frac{1}{2}(|w_i-w|/\lambda_*)^2}} {(2\pi\lambda_*^2)^{1/2}}}
\end{eqnarray}
where $w_i$ $(i=1\dots m)$ are the scalar values of the randomizations, and $\lambda_*$ is a 
real-valued smoothing parameter. By partitioning the data into 
five ``folds'' and holding out one fold at a time to calculate the 
average probability of a hold-out set according to the other 
4/5 of the data (``5-fold cross-validation''~\cite{Hastie}),
we define the function
\begin{equation} 
Q(\lambda)\equiv\frac{1}{5}\sum_{i=1}^5\Pi_{j=1}^{\frac{4}{5}m}{{p_\lambda(w_{f_i(j)})}}.
\end{equation}
where $\{f_i(j)\}_j$ is the set of
indices associated with fold $i$ ($i=1\dots5$) 
We then determine $\lambda_*$ as \mbox{$\lambda_*\equiv\mbox{argmax}_\lambda{Q(\lambda)}$}.
For a real-world graph of interest, ranking of likelihoods
reveals the most significant measures of the network--the scalars which are least like the background ensemble. Figures \ref{kde1} and \ref{kde2} show the results of density estimation on the two most significant scalars.\\
{\bf{Localization}:}
Consider the set of scalars for digraphs, 
$\sum(B_1B_2\dots B_n)$ ($B_i\in\{A,A^T\}$,  $n\in\mathbb{N}$). These scalars perform a census which includes all possible walks and therefore all possible subgraphs. The operators $D$ and $U$ constrain the set of all subgraphs so that a given scalar only counts a small subset simultaneously. In this way scalars inherit statistical significance from subgraphs. While some scalars count an individual subgraph, other scalars count combinations of subgraphs. The mapping of scalars to subgraphs is thus many-to-many. \\
While the analysis proceeds independently of subgraphs, it 
is possible, given a graph, to find any scalar's  most
representative set of subgraphs. We call this process {\it localization}. We define a skeleton to be the smallest
subgraph with nonzero value of the scalar. 
As an approximate, greedy
algorithm to find a most representative set of skeletons, 
given a graph $A$ with nonzero value of a scalar $\cal W$, we: 
\begin{enumerate}
\item
build a subgraph, $s$, by adding nodes from $A$ until $\cal W$ 
evaluated on $s$ gives a nonzero value (soft-localization) or the original value (hard-localization);  
\item
 {\it distill} this subgraph by removing nodes from
$s$ until we arrive at a subgraph, 
$s'$, such that removing any additional nodes would cause the value of
$\cal W$ to vanish;
\item
repeat on $A-s'$ until all nodes have been exhausted.
\end{enumerate}
The resulting algorithm yields a set of representative subgraphs for a given scalar. Each $s'$ subgraph in the set is labeled according to its isomorphism class.  The most representative subgraph is simply the subgraph class which has the highest relative fraction of the total recovered set of subgraphs. Multiple iterations of the localization algorithm should be run since the algorithm depends on the order of the nodes; however, in practice, we do not see differences in results using different orderings. \\
As an example, hard and soft localization of the 
scalar $\sum(DAUA^TA)$ from the \emph{E. coli} genetic 
network
reveals the $\cal {T}$$_{030}$ triad (Figure \ref{fflsimdor}a)
familiar from \cite{Alon1}.
Arbitrarily large substructures may 
emerge from a given scalar,
highlighting 
another 
methodological advantage to the algorithm: 
the search for significant scalars does not 
impose any constraints regarding the size of resulting subgraphs. An upper bound on the computational complexity is $(\ell-1)\sum_i^si^3$, where $s$ is the size of the resulting substructure and $\ell$ is the length of the scalar. In general, however, the efficiency of the localization algorithm is of less concern, as we localize only on a small set of statistically significant scalars.  \\
{\bf{\emph{E. coli} Data Set}:}
We implemented our algorithm
on the \emph{E. coli} genetic network. The database includes $577$ interactions between $423$ nodes, combining an existing database \cite{regulondb} with additional nodes and edges included from a literature search as described by Alon \emph{et al.} \cite{Alon1}. We exclude self-interactions for a total of $519$ edges.  
Density estimations (see Figures \ref{kde1}, \ref{kde2}, \ref{kde3}) demonstrate how \emph{E. coli} deviates from our background ensemble.
Three of the top-ranking statistically significant scalars, $\sum(DAUA^TA)$, $\sum(DA^TAUA^TA)$, 
and $\sum(DA^TAA^TA)$, localize to several structures consistent with Shen-Orr et al.'s earlier findings with this data set (see Figure \ref{fflsimdor}). However, we highlight that identification of these three significant structures was done using one algorithm without the need to pose thresholds or parameters or to provide tailored algorithms. No property of the network was assumed to be of interest beforehand. Of interest, $\sum(AA^TDA^TAA)$ was the highest scoring scalar.  Upon soft-localization, we recovered 
the two 4-node subgraphs (Figure \ref{soft1}), which we call ``FFB" (feed-forward box) and ``+FFL" (feed-forward loop with an input). The 4-node structures are more significant than the related 3-node $\cal {T}$$_{030}$ topology. The methodology thus assigns significance to a scalar without biasing the size of the resulting subgraphs. \\ 
Closer inspection of the top-scoring scalars reveals some unexpected architectural features. Hard-localization of the significant scalar, $\sum(AA^TDA^TAA)$, yields a 14-node topology (Figure \ref{hard1}). We observe that the $\cal {T}$$_{030}$ topology, defined by the genes $hns$, $flhDC$, and $fliA$, is a motif shared by five overlapping FFB's. Inspecting the word, $DA^TAA$, on the \emph{E. coli}
data, we find that there are $42$ distinct $\cal {T}$$_{030}$ paths, but only $10$
distinct $\cal {T}$$_{030}$ grandparents. That is, the operation $\sum$
evaluates to $42$, while the operation $\cal N$ evaluates to  
$10$.  In fact, the gene $crp$ appears in $16$
`distinct' $\cal {T}$$_{030}$. In this way the network evidences {\it motif hubs}--individual nodes which appear in numerous, overlapping identical motifs, a result first noted in \cite{matstat} using a more primitive significance test and more recently reported in \cite{Dobrin}. Importantly, this result is obtained with a single algorithm without posing any prior assumptions about the network.  \\
Scalars which are significantly smaller relative to the background ensemble also reveal interesting topological features of the graph. For example, we find the scalar $\cal{N}$$(UA^TA)$ is statistically underrepresented in the \emph{E. coli} network (see Figure \ref{kde3}). Localizations reveal structures with nodes that have two or more incoming edges. This ``fan-in" structure, the opposite of the ``SIM" topology, thus appears less often in the network, a finding with important ramifications. For example, recently researchers attempting to infer genetic regulatory interactions have imposed priors which restrict the number of edges converging on a node, but leave unrestricted the number of edges leaving a node \cite{Husmeier}. This prior on a general ``fan-out" topology is thus supported by our findings.   \\
{\bf{\emph{S. cerevisiae} Data Set}:}
The yeast dataset is based on the Yeast Proteome Database (YPD) \cite{ypd} and this particular part of the network consists of $688$ nodes with $1079$ edges \cite{yeastsite}. Analysis of this network shows 
the most significant word, $\sum(DAAA^TDAA)$, 
contains a mutual dyad (a term which we borrow from the sociological network literature, referring to a pair of vertices mutually linked, such that, $a \rightleftarrows b$) as the rightmost $DAA$ indicates.  Upon 
hard-localization we find that only four nodes in the network contribute to the word;
these four nodes make up a dense cluster 
which includes a mutual dyad and a 3-cycle (see Figure \ref{m3c}). 
Another significant feature, $\sum(DAAAUA^TA)$, 
hard-localizes to a 22-node substructure (Figure \ref{yeast22})
with a  
fascinating topology which includes two 
parent genes which have a large and almost identical set of children.  
In the soft-localization of this feature,  
a minimal subgraph
emerges with a compound topology: the ``parent" layer of an FFL is itself 
an FFL 
(Figure \ref{soft2}). Obviously this $5$-node subgraph would not be identified with subgraph census methods which only count up to $3$- and $4$- node subgraphs. \\
{\bf{Interpretation}:}
We have presented a generalizable method for enumerating measures of a network and have demonstrated an application of this method for finding statistically significant features of the \emph{E. coli} and \emph{S. cerevisiae} genetic regulatory networks. The method has the advantages of \emph{computational efficiency} as compared to subgraph census for naturally-occurring networks, particularly clustered or scale-free networks, and \emph{flexibility} in that it can be easily applied to weighted and signed graphs. For example, many biological networks are published with a ``p value" associated with each edge \cite{olga,droso}, i.e., a probability that a certain edge exists (implicit in such publications is the assumption that the existence of each edge is independent of all other edges). In this case, $\Sigma$ refers to the expected value of that scalar, over all realizations of the graph. Alternatively, neuronal networks have weighted edges describing the number of synapses and thus the strength of the interaction. In this case, $\Sigma$ calculates the functionality of a particular word. 
While the $\cal {N}$ operation does not differentiate between weighted and unweighted edges, we could easily imagine other useful quantities of interest that we can also use in our space that would be functions of the weighted edges, such as the standard deviation. 
Some of our results on \emph{E. coli} confirm earlier findings from previous methods, but unlike those methods, this approach is a single, systematic algorithm which does not require any previous assumptions about the network being analyzed. Moreover, new results regarding the structure of the \emph{E. coli} network are presented, including the presence of ``motif-hubs," ``feed-forward boxes" and a general ``fan-out" topology. \\
It is worth highlighting that under a different randomization 
scheme 
with a different set of 
conditionals, the results may differ substantially.
For example, in the case of the yeast dataset, if the number of 3-cyles or mutual dyads was also preserved, we
expect the ranking of scalars to be different. 
We note, then, that one must take great care in selecting
the background ensemble
to avoid the possibility that one's choice of 
randomization 
predetermines which 
scalars
are
the 
most significant. While the configuration model and its variants have been used as the appropriate ensemble distribution for networks in the past, many other random network models exist which may be more appropriate. Potentially, the network embedding space we present here will elucidate these issues further. For example, given multiple realizations of two random network models, one can use this space to investigate whether the resulting distributions are separable and which features make them distinguishable.\\
While motivated by work in which the subgraphs are the primitive
degrees of freedom, scalars do not have a one-to-one mapping to 
subgraphs.
However, every subgraph contributes to at least
one scalar. 
Subgraph counting is computationally expensive, particularly for clustered, scale-free, and dense networks, but our method alleviates this issue because its exponential term is independent of these properties. The trade-off is that with localization, we can only find sets of subgraphs that a given scalar counts. A more systematic alphabet could further constrain the set of subgraphs for a scalar. \\
Closer investigation into the mapping between scalars and subgraphs is needed.
The heuristic we develop, localization, appears to work well. The scalars are easily mapped to their most representative subgraphs, and some of these subgraphs confirmed earlier findings on the same dataset. However, while in our studies the interpretation of the most representative subgraphs of the significant scalars was straightforward, some scalars may have more difficult interpretations. We note that our focus here was not on subgraphs per se, but rather on a data space, a set of measures on a graph from which one can perform various statistical studies. In general, if one is interested in a particular subgraph, then the best approach is to identify that subgraph in the network. If one does not have any preconceptions about which feature of the network is important to study, than the scalar space offers an alternative, systematic, efficient, and effective approach to census and/or listing properties deemed relevant. Indeed, the space may not only be related to subgraphs, but also to more global measures such as various orders of transitivity \cite{newman}. \\ 
Finally we note 
additional utilities of the enumeration of words.
First, given an algorithm which purports to model a real-world network, one could find statistically
significant scalars to identify in what ways the model fails to model the real-world data.
Second, given a
training set of many graphs of multiple classes, this 
data
space
could be used to build a classifier using machine learning
algorithms ({\it e.g.}, SVMs \cite{Vapnik,Cristianini}, Boosting \cite{Freund})
which could then assign new graphs to one of the classes (see \cite{netclass,droso} for recent work in this direction), providing a modern machine learning
approach for diagnosing networks ({\it e.g.}, 
robust vs fragile economies, graphs with different growth laws, etc.). \\
It is a pleasure to acknowledge useful conversations with
U. Alon,
C. Stein,
J. Gross,
R. Albert,
and
M. Newman.
We thank 
the organizers of the LANL/CNLS conference on 
``Networks: Structure, Dynamics, and Function" 5/03. CW was supported in 
part by NSF ECS-0332479, NSF DMS-9810750, NIH GM36277, NIH LM07276.

\bibliographystyle{plain}
\bibliography{matstat_arxiv}

\end{document}
\end